\documentclass[journal]{IEEEtran}
\ifCLASSINFOpdf
\usepackage{graphicx}
\usepackage{lipsum} 
\else
\fi
\usepackage{tabularx}
\usepackage{amssymb}
\usepackage{amsmath}
\usepackage{algorithmic}
\usepackage{etoolbox}
\usepackage{amsmath}
\usepackage{algorithm}
\usepackage[a4paper,margin=1in]{geometry}
\usepackage{fancyhdr}

\usepackage{array}
\newcolumntype{E}[1]{>{\raggedright\let\newline\\\arraybackslash\hspace{0pt}}m{#1}}
\newcolumntype{L}[1]{>{\raggedright\let\newline\\\arraybackslash\hspace{0pt}}m{#1}}
\newcolumntype{C}[1]{>{\raggedright\let\newline\\\arraybackslash\hspace{0pt}}m{#1}}
\newcolumntype{R}[1]{>{\raggedright\let\newline\\\arraybackslash\hspace{0pt}}m{#1}}
\makeatletter
\patchcmd{\@makecaption}
{\scshape}
{}
{}
{}
\makeatletter
\patchcmd{\@makecaption}
{\\}
{.\ }
{}
{}
\makeatother

\usepackage{xcolor}
\begin{document}

\title{\small{ \textcolor {blue}{This work has been submitted to the IEEE for possible publication. Copyright may be transferred without notice, after which this version may no longer be accessible.}} \\ \huge Connectivity and Collision Constrained Opportunistic Routing for Emergency Communication using UAV }
%
% author names and IEEE memberships
% note positions of commas and nonbreaking spaces ( ~ ) LaTeX will not break
% a structure at a ~ so this keeps an author's name from being broken across
% two lines.
% use \thanks{} to gain access to the first footnote area
% a separate \thanks must be used for each paragraph as LaTeX2e's \thanks
% was not built to handle multiple paragraphs
%

\author{Sharvari  NP,~\IEEEmembership{Student Member,~IEEE,}
        Dibakar Das,~\IEEEmembership{Member,~IEEE,}
        Jyotsna Bapat,~\IEEEmembership{Member,~IEEE}
        and Debabrata Das,~\IEEEmembership{Senior Member,~IEEE}
        \thanks{The authors are with  International Institute of Information Technology Bangalore (IIITB), Bengaluru, India.}% <-this % stops a space
%        \thanks{J. Doe and J. Doe are with Anonymous University.}% <-this % stops a space
        %\thanks{Manuscript received April 19, 2005; revised August 26, 2015.}
         % <-this % stops a space
%\thanks{M. Shell is with the Department
%of Electrical and Computer Engineering, Georgia Institute of %Technology, Atlanta,
%GA, 30332 USA e-mail: (see %http://www.michaelshell.org/contact.html).}% <-this % stops a space
%\thanks{J. Doe and J. Doe are with Anonymous University.}% <-this % stops a space
%\thanks{Manuscript received April 19, 2005; revised December 27, 2012.}
}

\maketitle

% As a general rule, do not put math, special symbols or citations
% in the abstract or keywords.
\begin{abstract}
Emergency communication is extremely important to aid rescue and search operation in the aftermath of any disaster. In such scenario, Unmanned Aerial Vehicle (UAV) networks may be used to complement the damaged cellular networks over large areas. However, in such UAV networks, routing is a challenge, owing to high UAV mobility, intermittent link quality between UAVs, dynamic three dimensional (3D) UAV topology and resource constraints. Though several UAV routing approaches have been proposed, none of them so far have addressed  inter UAV coverage, collision and routing in an integrated manner. In this paper, we consider a scenario
where network of UAVs, operating at different heights from ground, with inter UAV coverage and collision constraints, are sent on a mission to collect disaster surveillance data and route it to Terrestrial Base Station via multi-hop UAV path. Analytical expressions for coverage probability ($P_{cov}$) and collision probability ($P_{coll}$) are derived and minimum ($R_{min}$) and maximum ($R_{max}$) distance between UAVs are empirically calculated. We then propose a novel Multi-hop Opportunistic 3D Routing (MO3DR) algorithm with inter UAV coverage and collision constraints such that at every hop expected progress of data packet is maximized. The numerical results obtained from closed form mathematical modelling are validated through extensive simulation and their trade-off with variation in network parameters such as path loss component, trajectory divergence etc. are demonstrated. Finally, for application requirement of $P_{cov} \geq 0.8$ and $P_{coll} = 0$ we obtain empirical optimality condition for inter UAV distance as $R_{min} \geq 10$m and $R_{max} \leq 60$m.
\end{abstract}

% Note that keywords are not normally used for peerreview papers.
\begin{IEEEkeywords}
Unmanned Aerial Vehicles, emergency communication, opportunistic routing, coverage probability, collision probability.
\end{IEEEkeywords}

% For peer review papers, you can put extra information on the cover
% page as needed:
% \ifCLASSOPTIONpeerreview
% \begin{center} \bfseries EDICS Category: 3-BBND \end{center}
% \fi
%
% For peerreview papers, this IEEEtran command inserts a page break and
% creates the second title. It will be ignored for other modes.
\IEEEpeerreviewmaketitle

\section{Introduction}
% The very first letter is a 2 line initial drop letter followed
% by the rest of the first word in caps.
% 
% form to use if the first word consists of a single letter:
% \IEEEPARstart{A}{demo} file is ....
% 
% form to use if you need the single drop letter followed by
% normal text (unknown if ever used by IEEE):
% \IEEEPARstart{A}{}demo file is ....
% 
% Some journals put the first two words in caps:
% \IEEEPARstart{T}{his demo} file is ....
% 
% Here we have the typical use of a "T" for an initial drop letter
% and "HIS" in caps to complete the first word.
\IEEEPARstart{N}{atural}  disasters e.g. floods, etc., often yield annihilating consequences. The usual aftermath of such unexpected events is partial or complete destruction of existing terrestrial communication network. In these situations where existing cellular communication network are compromised, there comes a crucial requirement to revamp the wireless infrastructure to aid rescue and search operations. Consequently, a robust, fast, effective emergency surveillance and communication system is needed to enable reliable disaster related information dissemination. Unmanned Aerial Vehicles (UAVs) with their high mobility, path optimization, Line of Sight (LOS) communication  can be used for such applications. However, network of UAVs needs to be connected to distant Terrestrial Base Stations (TBS) to transfer critical data to ground control stations for rescue process. Thus, a combination of UAVs and TBS can offer a reliable communication for disaster relief operations.

Off late UAV assisted communication and networking have attracted much interest from both academia and industry, owing to their enormous potential in both civil and military domains [1][2]. UAVs come with several enhancements over traditional wireless infrastructure such as ability to intelligently self-adjust their positions in real time, ability to provide uninterrupted communication while moving at high speeds and ability to benefit from high altitude operations to provide unobstructed wireless channels in air. These advantages make UAVs a favourite candidate for myriad of applications like, mapping, surveillance, security, traffic control, package delivery etc. UAVs can be deployed as both aerial base stations  and flying user equipments [4][5][6][7][8][9][10][11][12]. 

For efficient cooperation and coordination between multiple UAVs, reliable  wireless communication between  UAVs communicating with each other and in addition with the TBS
is extremely important. This requires efficient routing protocol for reliable transmission of data between UAV nodes. Despite similarity of UAV networks (UAVNETS) with Mobile ad hoc network (MANETS) and Vehicular ad hoc network (VANETS), the applicability of traditional routing protocols and its variants have shown limited network performance owing to challenges associated with UAVs such as rapid mobility and highly dynamic topology. Table I summarizes some of the major differences that set UAVNETS apart from  MANETS and VANETS [13][14][15].

\begin{table}
	\centering
	\caption {Comparison between MANETS, VANETS and UAVNETS.}
	\begin{tabular}{| E{1cm} | L{1.5cm} | C{1.5cm} | R{1.5cm} |}
		\hline
		\textbf{Parameter} & \textbf{MANETS}& \textbf{VANETS}& \textbf{UAVNETS}\\
		\hline
		Specification & Ad hoc networks in which nodes are mobile, communicating with other nodes within range. & Vehicles are mobile nodes, communicating with each other and with the road side unit in ad hoc manner.& Aerial nodes/UAVs communicating with each other and with the ground/control station in ad hoc manner.  \\
		\hline
		Speed & Slow, usually around 2 m/s. & Comparatively high speed, around 20-30 m/s in 2 dimension. & Speed range from 0 m/s (in case of stationary operation) to 100 m/s in 2/3 dimension.\\
		\hline
		Topology Changes & Dynamic nodes  that join and leave the network intermittently. & More dynamic than MANETS. & Stationary, slow (in case of hovering operation) or fast depending on the mission.\\
		\hline
		Energy constraint & Battery powered that could last for few hours. & Devices may be charged from vehicles battery while in motion or own powered. & Comparatively more energy constraint, needs frequent charging.\\
		\hline
	
	\end{tabular}
\end{table}

Though several routing approaches such as   single hop routing, topology based routing, position based routing, cluster based routing etc. have been studied for surveillance application, yet there exist a lacunae when it comes to their applicability to UAV networks in terms of high mobility, resource constraints, sparse deployment and highly dynamic topology [16][17][18]. Especially when UAVs are deployed for disaster surveillance application with intermittent links and nodes, they present unique set of routing challenges. First, the availability of receiver node within the coverage range of a transmitting node is uncertain. Second, even before the transmission of the data it needs to be ascertained whether the transmitting node and the receiving peer  are likely to collide, leading to loss of communication. Third, there could be range restrictions between UAVs and TBS, often giving rise to the requirement of multi-hop communication between UAVs and TBS. Consequently, it needs to be determined which peer node within the coverage range of the transmitting node is likely to transmit the data in the direction of the destination. The above challenges in achieving reliable UAV communication has not been addressed in an integrated manner in any of the previous work to the best of our knowledge. 

\begin{figure}[!t]
	\centering
	\includegraphics[width=2.5 in]{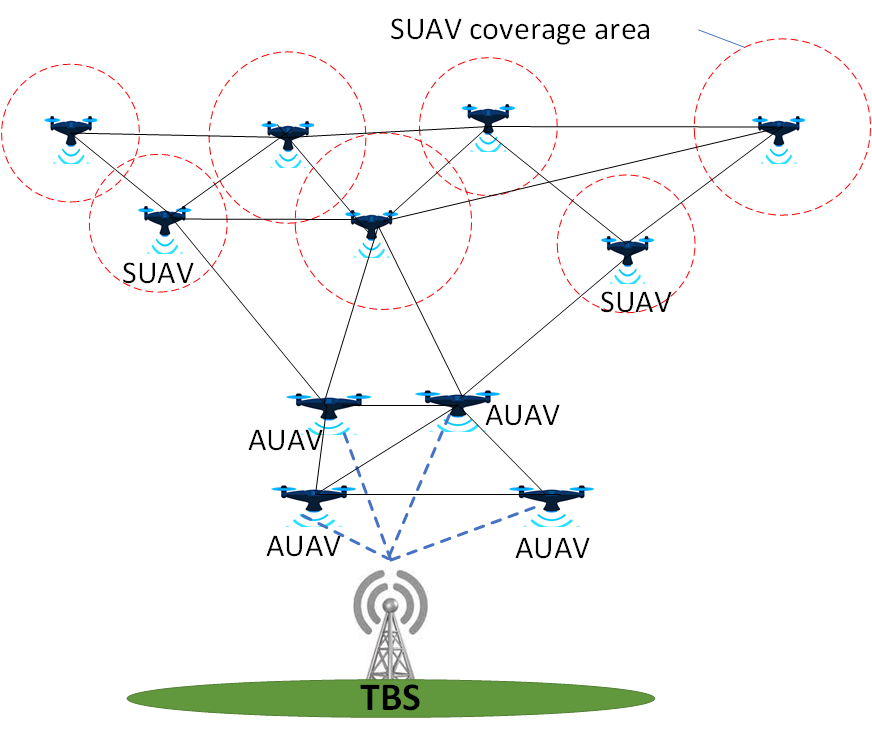}
	% where an .eps filename suffix will be assumed under latex, 
	% and a .pdf suffix will be assumed for pdflatex; or what has been declared
	% via \DeclareGraphicsExtensions.
	\caption{Illustration of UAV enabled communication network. (SUAV - Surveillance UAV, AUAV - Anchored UAV and TBS - Terrestrial Base Station).}
	%\label{fig_sim}
\end{figure}
In this paper, we investigate UAV routing design problem in a UAV enabled communication system for disaster surveillance, with multiple UAVs communicating with each other and with the remote TBS. As depicted in Fig.1 the UAVs have a mission of flying from their initial location to explore the disaster affected areas, to gather information on the impact of the event and then route the collected data to TBS to aid rescue and search operations. The UAV nodes are linked with each other and with the TBS. Here the communication can take place in two patterns, UAV to UAV and UAV to TBS. To enable such a communication the UAVs are grouped as Anchored UAVs (AUAV) and Surveillance UAVs (SUAV). The AUAVs are positioned at specific locations within the coverage range of TBS, avoiding any obstacle, to enable uninterrupted connection to TBS for data dissemination; while the SUAVs are responsible for flying to destination locations for data collection. We then propose a novel Multi-hop Opportunistic 3D Routing (MO3DR) algorithm under connectivity and collision constraints to facilitate inter UAV communication. 

The contributions of this paper can be stated as follows, 
\begin{enumerate}
\item We provide analytical expressions for  probability of coverage ($P_{cov}$) and  probability of collision  ($P_{coll}$), for a typical multi-UAV network. The proposed model considers various parameters such as propagation channel characteristics, UAV network distribution density, UAV flying altitude, UAV trajectory divergence etc. $P_{cov}$ and $P_{coll}$ represent the coverage and collision constraints respectively. The coverage constraint is given by the minimum Signal to Interference Ratio (SIR) requirement which needs to be satisfied at all time along the UAV path. This corresponds to the downlink payload data (real time video, photo) transmission from UAVs to TBS. Collision constraint is based on the position of the UAV and its neighbours so that they are operated at safe distance from each other to avoid collision. 

\item Using the obtained analytical expressions we establish that there exist an optimal range of inter UAV distance between two communicating UAVs. Lower bound is given by  $R_{min}$ and upper bound by $R_{max}$. $R_{min}$ is chosen such that $P_{coll}$ is minimized where as,  $R_{max}$ is defined such that $P_{cov}$ is maximized as per the requirements of the underlying application.

\item We obtain numerical  results using expressions derived for $P_{coll}$ and $P_{cov}$. Later, simulations are run to validate the proposed model. These results demonstrate that the coverage probability is highly dependent on the network parameters such as UAV node density, path loss component and number of UAV nodes while collision probability depends on the trajectory divergence.

\item We then propose a novel MO3DR algorithm with an objective that the data packet makes maximum progress towards the destination at every hop. To ensure this we select the next hop nodes within a specified sector. The effect of the sector angle on the expected progress for varying network density is mathematically analysed. Finally, the simulation results verify the validity and effectiveness of the proposed routing algorithm.
\end{enumerate}

The reminder of the paper is organized as follows. In section II, the existing related work on the applicability of UAVs for surveillance and monitoring services and UAV routing protocols are surveyed in detail. Section III describes the system model of UAV enabled cellular communication system for disaster surveillance. It comprises of mathematical derivations of probability of coverage, probability of collision and expected progress of packets towards destination as proposed by MO3DR algorithm. Section IV extensively discuss the results obtained from the closed form mathematical expressions and simulation. Finally, the paper is concluded in section V.

% You must have at least 2 lines in the paragraph with the drop letter
% (should never be an issue)
%I wish you the best of success.

%\hfill mds
 
%\hfill December 27, 2012
\section{Relevant Work}
Motivated by the potential applications of UAVs and new design challenges of UAV assisted wireless communication as discussed earlier, there has been an overwhelming interest in research in these directions. This section presents a detailed survey of topics related to this work.
\subsection{UAVs as Aerial Base Stations}
Here, we discuss the applicability of UAV mounted Aerial Base Stations (ABS) in disaster environment and monitoring to boost the performance of existing terrestrial wireless network with enhancement in coverage, delay and Quality of Service (QoS) [3]. In particular, the UAV ABS can be of great help for public safety communication for search and rescue operations. Various Public Safety System (PSS) based on Wi-Fi, 4G Long Term Evolution (LTE), satellite communication, dedicated PSS like TETRA have been extensively used [4]. But these technologies may not be dynamic enough to adapt to fast changing environment during disaster and may not be able to provide low latency communication. In this regard, UAV ABS can act as an alternative means of communication network during destruction of existing  infrastructure in the aftermath of disaster and help rescue operation. The placement of UAVs become at most important when used as ABS to enhance coverage and meet energy efficiency constraints [5][6]. The UAV ABS have been  studied for supplementing existing infrastructure, to act as relay between transmitter and receiver providing intermediary coverage and to enhance connectivity  to the terrestrial network [7]. 

\subsection{UAVs as User Equipment}
UAV as flying User Equipment (UE) are deployed for package delivery, surveillance, monitoring, remote sensing etc. [8]. With the advantage of ability to quickly optimize path as per the changing dynamics of environment or network conditions, UAVs are extensively studied to be deployed as cellular connected UEs. The UAVs are majorly used for information dissemination, because of their high mobility and LoS opportunity. They also have been successfully used as relays to maximize communication coverage range, on demand connectivity, traffic offloading and information dissemination to ground stations [9]. Another important application of UAV UE is virtual reality (VR), where the data captured by the UAVs are transmitted to remote VR users [10][11]. However, the UAV UEs are significantly different from that of terrestrial UEs, mainly due to channel conditions which is predominantly LoS, limited onboard energy and more dynamic as they can continuously fly without much restriction like on ground mobility [12]. Therefore, the incorporation of UAV UE introduce new design considerations that needs to be addressed.

\subsection{Inter UAV routing }
In addition to the requirements in conventional wireless networks, such as allowing the network to scale, finding optimal route, meeting latency constraints, reliability and required QoS; UAV networks should also consider stringent energy constraints, location awareness, should be adaptable to intermittent links, frequent add/drop of UAV nodes and dynamic topology [13]. In this regard several works have investigated the use of existing routing protocols and their variants including static routing protocols, proactive routing protocol, reactive routing protocols and geographic 3D routing protocols for the possible use in UAV networks [14]. Various static routing protocols like Load, Carry and Forward Routing Protocol, Multi-Level Hierarchy Routing Protocol, Data Centric Routing Protocol have been studied for UAV application. But since the routing here is based on static routing tables that are computed and loaded when the tasks start make them not suitable for dynamically changing environment [15]. Proactive Routing Protocols like Optimized Link State Routing, Destination Sequenced Distance Vector Routing have also been studied. But huge message overhead that are exchanged between the nodes to keep the routing table updated make them unsuitable for UAV network because of bandwidth, energy, and delay constraints [16]. Reactive Routing Protocols like Dynamic Source Routing, Ad-hoc on-Demand Distance Vector Routing  have proved to be incompetent due to scalability issues [17]. On the other hand, Geographic 3D Routing Protocols like Greedy Hull Greedy, the locations may not get updated as rapidly as topology changes. Hence make them incompetent to cater the needs of the UAV routing [18]. In this regard, various dimensions like deterministic, stochastic and opportunistic are being analysed to design UAV routing protocols [19][20]. 

\section{System Model}	
As shown in Fig. 1, we consider UAV enabled  communication system with $M$ UAVs and one TBS. The set of $M$ UAVs is denoted as $\psi_U$. The set $\psi_U$ is partitioned as sets $\psi_A$ and $\psi_S=\psi_U \setminus \psi_A$ representing AUAVs and all the remaining UAVs acting as SUAVs, respectively. The UAVs (both SUAV and AUAV) fly from their initial location $L_0 \in R^3$ to final location $L_F \in R^3$. The final location of AUAVs is fixed at specific positions within the coverage range of the TBS. The location of SUAVs are random in the given surveillance area. The initial location for all the UAVs is typically at the origin of the Euclidean plane represented as $(x_0 ,y_0 ,h_0)$. The coordinates of $m^{th}$ UAV is denoted as $(x_m,y_m,h_m )$ as shown in Fig. 2. The parameters used in the analytical modelling are presented in Table II.
\begin{figure}[!t]
	\centering
	\includegraphics[width=2.5 in]{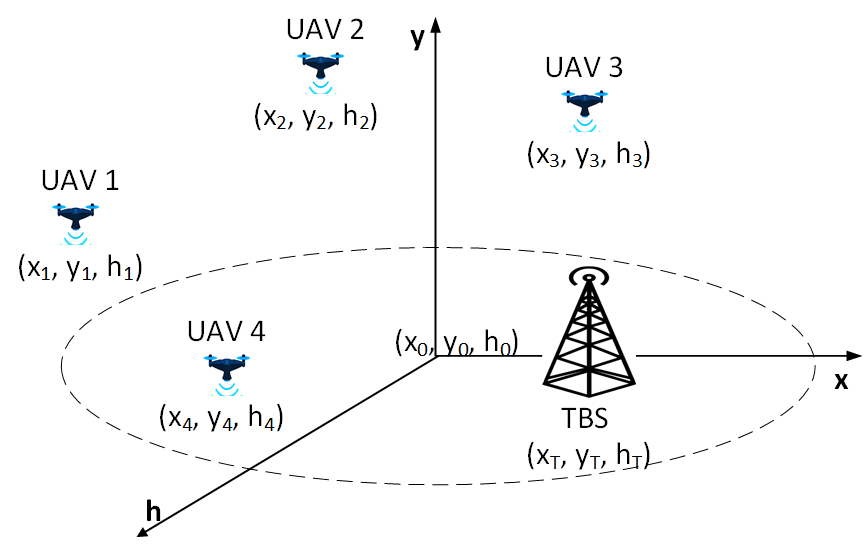}
	% where an .eps filename suffix will be assumed under latex, 
	% and a .pdf suffix will be assumed for pdflatex; or what has been declared
	% via \DeclareGraphicsExtensions.
	\caption{Illustration of the initial and final locations of UAVs and  location of TBS in 3D space.}
	%\label{fig_sim}
\end{figure}
\subsection{Propagation model}
For the purpose of elucidation, we assume that the UAVs and TBS are equipped with a single antenna with omnidirectional radiation pattern. Channel for inter UAV and UAV-TBS communication is assumed to be predominantly LoS link with propagating signal subjected to distance dependent pathloss and small-scale fading. Standard path loss model is adopted and pathloss function is defined as,
\begin{equation}
l(r_{ij}, h_i,\alpha )= (r_{ij}^2 + h_i^2)^{-\alpha/2}
\end{equation}
where, $r_{ij}$   is the distance between the $i^{th}$ transmitting UAV and $j^{th}$ receiving UAV, $h_i $ is the altitude from ground at which $i^{th}$ UAV is operated and $\alpha$ is the path loss exponent.  We assume Nakagami-$m$ small scale fading channel  with multipath fading component $H_{ij}$ with fading parameter $\delta$. Nakagami-$m$ small scale fading channel is presumed as this is the generalised fading model that can represent a range of wireless scenarios.

The Signal to Interference Noise Ratio (SINR) at the receiving UAV in  $\psi_U$ is described as,
\begin{equation}
SINR = \frac{S_*}{{I + \sigma}^2}
\end{equation}
where, $S_*$ is the signal power received from the transmitting UAV and $I$ is the interference signal defined as, the aggregate signal power received from all UAVs in $\psi_U$ other than the transmitting UAV and $\sigma^2$   is the Gaussian distributed noise power. We assume interference limited scenario where $I \gg \sigma^2 $ [21]  and hence consider Signal to Interference Ratio (SIR) as,
\begin{equation}
SIR =\frac{S_*}{I }
\end{equation}
Following the channel model in [22], the instantaneous received signal power at $j^{th}$ UAV from $i^{th}$ UAV is defined as,
\begin{equation}
S_{ji} = H_{ij} l(r_{ij}, h_i, \alpha)
\end{equation}

\begin{table}
	\centering
	\caption {List of notations used in analytical modelling.}
	\begin{tabular}{l l}
		\hline
		%\hline\\
		\textbf{symbol} & \textbf{specification}\\
		\hline
		%\hline\\
		$P_{cov}$ & probability of coverage \\
		$P_{coll}$ & probability of collision \\
		$\psi_{U}$ & set of all UAVs  \\
		$\psi_{A}$ & set of Anchored UAVs \\
		$\psi_{S}$ &set of Surveillance UAVs \\
		$L_{0}$ & UAV initial location   \\
		$L_{F}$ & UAV final location   \\
		$S_{*}$ & received signal power	\\
		I & interference signal	power	\\
		$\sigma$ & Gaussian distributed noise power\\
		$h_{*}$ & height above the ground\\
   	    $r_{*}$ & inter UAV distance	\\
   	    $H_*$ & small scale multipath fading component	\\
	    $\delta$ & channel dependent fading parameter	\\
	    $\alpha$ & path loss exponent\\
	    $\xi$ & trajectory divergence	\\
		$\phi$  & sector angle \\
		$R_{max}$ & maximum inter UAV distance\\
		$R_{min}$ & minimum inter UAV distance	\\
		\hline		
	\end{tabular}
\end{table}

\subsection{Inter UAV distance distribution }
To analytically characterise the inter UAV distance, it is assumed as $m$ dimensional Poisson Point Process (PPP) of intensity $\lambda$. In this scenario, the usage of PPP allows to capture the spatial randomness of UAV nodes in practical networks and at the same time obtain compliant closed form expressions for system level performance matrices [23]. The inter UAV distance is denoted by a random variable $R_n$. The distance between  UAV and its $n^{th}$ neighbour is distributed according to generalized Gamma distribution following the distance distribution model in [24]. The Probability Distribution Function (PDF) of  $R_n$ is given by, 
\begin{equation}
f_{R_n} (r)=exp(-\lambda v_m r^m) \frac{m{(-\lambda v_m r^m)}^n}{r\Gamma(n)} 
\end{equation}
where, $v_m r^m$ is the volume of $m$ dimensional ball of radius $r$. Cumulative Distribution Function (CDF) of  $R_n$ is given by,
\begin{equation}
F_{R_n} (r)=1-\frac{\Gamma_*(n, \lambda v_m r^m)}{\Gamma(n)}
\end{equation}
where, $\Gamma_*$ is the incomplete Gamma function. 
\subsection{ Probability of coverage}
For successful data dissemination between the SUAVs and SUAV-AUAV each node must be able to communicate with every other node via a multi-hop path. To make sure that the communication between any two UAVs is not interrupted, connection is established between only those UAVs within the permissible range of inter UAV distance  $R_{min}  \leq r \leq R_{max}$. We derive the analytical expressions for $P_{cov}$ and then correspondingly deduce $R_{max}$. The parameter $R_{max}$ specifies the maximum allowable inter UAV separation such that the receiver UAV is within the coverage range of the transmitting UAV. $P_{cov}$ is defined as the probability that the SIR of received signal at the receptor UAV is above the threshold $\theta$. It is expressed as,
\begin{equation}
P_{cov}\triangleq P[SIR>\theta]
\end{equation}
The mathematical expression for $P_{cov}$ is defined as a function of interference signal $I$ and inter UAV distance $r$.\\\\
\textbf{Theorem 1:} The Probability of coverage is given by, 
\begin{equation}
P_{cov}= \int_{0}^{\infty}\sum\limits_{k=0}^{\delta-1} \frac{{(-1)}^k s^k}{k!}\bigg( \frac{d^k}{ds} \mathcal{L}_I (s)\bigg)f_{{R}_{n}}(r)dr
\end{equation}
where, $\mathcal{L}_I (s)$ is the Laplace transform of interference I.\\\\
\textbf{Proof:}The Probability of Coverage is derived as,

\begin{flalign} 
&\mathbb{P}[SIR \geq \theta] &
\end{flalign}
Substituting from (3) for $SIR$  we obtain the following expression,
\begin{flalign}
&= \mathbb{P}\bigg[\frac{S_*}{I} \geq \theta\bigg] &
\end{flalign}
Substituting from (4) for instantaneous received signal power at $j^{th}$ UAV from $i^{th}$ UAV we arrive at (11),
\begin{flalign}
&= \mathbb{P}\bigg[\frac{\mathcal{H}_{ij} {l(r_{ij},h_i,\alpha)}}{I} \geq \theta\bigg] &
\end{flalign}
From the definition of pathloss function $l(r_{ij},h_i)^{-\alpha/2} = (r_{ij}^2 + h_i^2)^{-\alpha/2} $  we obtain (12),
\begin{flalign}
&= \mathbb{P}\bigg[ \frac{\mathcal{H}_{ij} {(r_{ij}^2 + h_i^2)}^{-\alpha/2}}{I} \geq \theta\bigg] &
\end{flalign}
Rearranging the terms in (12) we get the following expression,
\begin{flalign}
&= \mathbb{P}\bigg[\mathcal{H}_{ij} \geq \frac{\theta I}{(r_{ij}^2 + h_i^2)^{-\alpha/2}}\bigg] &
\end{flalign}
\begin{flalign}
&= \int_{0}^{\infty}\mathbb{P}\bigg[\mathcal{H}_{ij} \geq \frac{\theta I}{(r_{ij}^2 + h_i^2)^{-\alpha/2}}\bigg]f_{{R}_{n}}(r)dr &
\end{flalign}
Since we have assumed Nakagami $ m $ - small scale fading channel, the random small-scale fading component $\mathcal{H}_{ij}$ is Gamma distributed random variable with channel dependent fading parameter $\delta$ [25]. Using this property, we arrive at (15),
\begin{flalign}
&= \int_{0}^{\infty}\mathbb{E}_I \Biggr[ \frac{\Gamma \bigg(\delta, \frac{\theta \delta I}{(r_{ij}^2 + h_i^2)^{-\alpha/2}}\bigg)}{\Gamma(\delta)} \Biggr]f_{{R}_{n}}(r)dr& 
\end{flalign}
Making the following substitution, $s=\frac{\theta \delta}{(r_{ij}^2 + h_i^2)^{-\alpha/2}}$ we get (16), 
\begin{flalign}
&= \int_{0}^{\infty}\mathbb{E}_I \bigg[\frac{\Gamma(\delta, sI)}{\Gamma(\delta)}\bigg]f_{{R}_{n}}(r)dr&
\end{flalign}
 Further the gamma function in (16) is expressed as $\Gamma(1+n,x)=n!exp(-x)\sum_{m=0}^{n}\frac{x^m}{m!}$ for $[n=0,1,\dots]$ [27] to get (17),
\begin{flalign}
&=\int_{0}^{\infty}\mathbb{E}_I \bigg[exp(-sI)\sum\limits_{k=0}^{\delta-1}\frac{(sI)^k}{k!}\bigg] f_{{R}_{n}}(r)dr&
\end{flalign}
 Using the substitution, $\frac{d^k}{ds}exp(-sI)=(-1)^k(I)^kexp(-sI)$ we arrive at (18),
\begin{flalign}
&= \int_{0}^{\infty}\sum\limits_{k=0}^{\delta-1}\frac{(-1)^k s^k}{k!}\mathbb{E}_I \bigg[\frac{d^k}{ds}exp(-sI)\bigg]f_{{R}_{n}}(r)dr&
\end{flalign}
Finally, (19) is obtained from Laplace Transform of the interference signal (proof in Appendix A).
\begin{flalign}
&=\int_{0}^{\infty}\sum\limits_{k=0}^{\delta-1}\frac{(-1)^k s^k}{k!} \bigg(\frac{d^k}{ds}\mathcal{L}_I(s)\bigg)f_{{R}_{n}}(r)dr&
\end{flalign}

\subsection{Probability of collision}
The lower bound of inter UAV distance  $R_{min}$ allows for collision free UAV movement. $P_{coll}$ is defined as the probability that two UAVs, reference and its neighbour will collide when the inter UAV distance between them is less than the sum of their radii, $r \leq r_R+ r_N$ as illustrated in Fig 3. The overall shape of the UAVs are assumed to be spherical, with radius of the reference UAV being $r_R$ and that of neighbour $r_N$. 
\begin{figure}[!t]
	\centering
	\includegraphics[width=2.5 in]{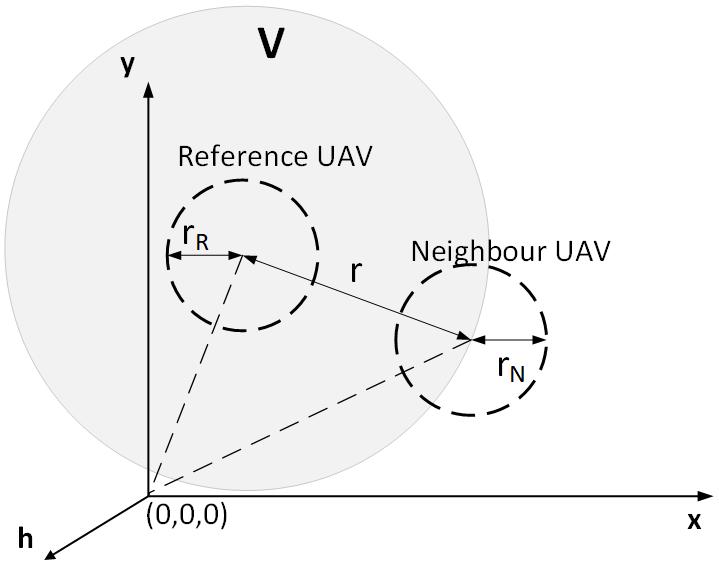}
%	% where an .eps filename suffix will be assumed under latex, 
%	% and a .pdf suffix will be assumed for pdflatex; or what has been declared
%	% via \DeclareGraphicsExtensions.
	\caption{Illustration of relative position of reference UAV and its neighbour.}
	%\label{fig_sim}
\end{figure}
The mathematical expression for $P_{coll}$ is derived as a function of inter UAV distance $r$ and trajectory divergence $\xi$. Trajectory divergence is defined as the deviation (in meters) in the UAV movement from its actual path.\\\\
\textbf{Theorem 2:} The Probability of collision is given by,
\begin{equation}
P_{coll} = exp\bigg(-\frac{r^2}{2\xi^2}\bigg)
\end{equation}
\textbf{Proof:} The computation of $P_{coll}$ is based on the presumption that the relative trajectory of UAVs is linear with fixed velocity. There exists positional errors due to trajectory divergence. These position uncertainties between any two UAVs (reference and its neighbour) is represented as 3D Gaussian distribution [26] with PDF $g(x,y,h)$. $P_{coll}$ is estimated as an integral of Gaussian PDF over 3D encounter space ($V$) as,
\begin{equation}
P_{coll}= \iiint\displaylimits_V g(x,y,h) \,dx\,dy\,dh
\end{equation}
where, $V$ is defined as the volume enclosed by sphere of radius $r$ within which the reference UAV is likely to collide with its neighbour. 

We assume the relative position variation along $h$ is very less compared to $x,y$.  Thus, the problem of $P_{coll}$ computation is transformed into a 2D integration of PDF. Hence, $g(x,y,h)$ gets reduced to (22),
\begin{equation}
g(x,y)= \frac{1}{2\pi \xi_x \xi_y}exp\Big[-\frac{1}{2} \Big( \frac{(x - \mu_x)^2}{\xi_x^2}+ \frac{(y - \mu_y)^2}{\xi_y^2}\Big)\Big]
\end{equation}
We assume that the coordinate system for the calculation of $P_{coll}$  is centered at the reference UAV. Thus, the 3D encounter space reduces to a circular cross sectional area of radius $r$ centered at $(0,0)$ as seen in Fig. 4; hence $\mu_x = \mu_y = 0$. Assuming uniform positional uncertainty along both $x$ and $y$ direction we get equi-trajectory divergent system with $\xi_x = \xi_y = \xi$.
$P_{coll}$ equals to the integration of PDF over the circular encounter cross sectional area as elucidated in (23)
\begin{figure}[!t]
	\centering
	\includegraphics[width=2.5 in]{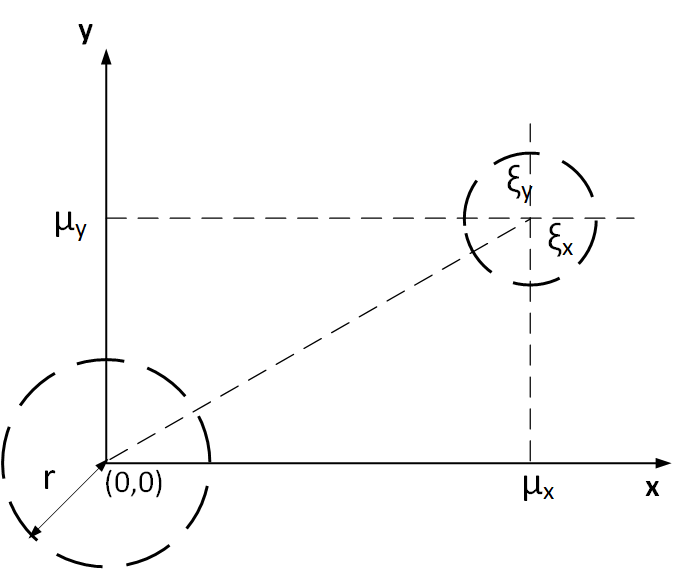}
%	% where an .eps filename suffix will be assumed under latex, 
%	% and a .pdf suffix will be assumed for pdflatex; or what has been declared
%	% via \DeclareGraphicsExtensions.
	\caption{Coordinate system for calculation of integral.}
	%\label{fig_sim}
\end{figure}
\begin{equation}
P_{coll}= \iint\displaylimits_{x^2+y^2\leq r^2} g(x,y) \,dx\,dy
\end{equation}
Substituting for $g(x,y)$ from (22) we get,
\begin{equation}
P_{coll}= \iint\displaylimits_{x^2+y^2\leq r^2} \frac{1}{2\pi \xi_2}exp\Big[-\frac{1}{2} \Big( \frac{x^2+y^2}{\xi^2}\Big)\Big]\,dx\,dy
\end{equation}
Transforming (24) into spherical coordinates with $x=rcos\theta$ and $y=rsin\theta$ we obtain (25), 
\begin{equation}
P_{coll} = \int\displaylimits_{r=0}^{r} \int\displaylimits_{\theta=0}^{2\pi} \frac{1}{2\pi \xi_2} exp \Big[-\frac{1}{2} \Big( \frac{r^2}{\xi^2}\Big)\Big] \,rdr\,d\theta
\end{equation}
On solving the integration in (25) we arrive at the final equation for $P_{coll}$ as a function of inter-UAV distance $r$ and trajectory divergence $\xi$ as,
\begin{equation}
P_{coll}= exp\bigg(\frac{-r^2}{2 \xi^2}\bigg)
\end{equation}

\subsection{Design of MO3DR Algorithm}
The proposed routing algorithm opportunistically forwards the data packets such that at every hop it maximizes the expected progress of the packet towards the destination. 

Consider a scenario with $M$ nodes (i.e. the total number of UAV nodes $= M$) with PPP distributed inter nodal distance. Here, several packets from SUAVs must be transmitted to the nearest AUAV via a multi-hop path. So, we determine the set of nodes that lie within the sector angle $\phi = (0 \leq \phi \leq \pi)$  that would be along $\pm {\frac{\phi}{2}} $ around $n_i-D$ axis (transmitting node-destination node axis; referred as direct path in the paper) as illustrated in Fig. 5. However, choosing an appropriate $\phi$ is very important as larger the $\phi$   larger could be the divergence from the direct path; smaller $\phi$  may result in very few/no potential receptor nodes. While choosing the next hop node we make sure that coverage and collision constraints are satisfied as per the requirements of the underlying application ($P_{cov} \geq P_{cov}^*,  P_{coll} \le P_{coll}^* $; $P_{cov}^*$ and $P_{coll}^*$ indicating application requirement). Among the nodes that satisfy the coverage and collision constraints, we then select the node that is nearest to $D$ with least divergence from direct path. This way we maximize the progress towards the destination as seen in Fig. 5.
The expected progress of the packet is obtained as, 
\begin{equation}
\mathbb{E}[R_n] = \int_{v_m r^m}rf_{R_n}(r)dr
\end{equation}
Substituting for  $f_{R_n}(r)$ from (5) we get,
\begin{equation}
\mathbb{E}[R_n]=\int_{v_m r^m}re^{-\lambda v_m r^m} \frac{m{(-\lambda v_m r^m)}^n}{r\Gamma(n)}dr
\end{equation}
Solving the above integral using the standard Gamma derivative identity $\int\displaylimits_{0}^{\infty}x^nexp(-ax)dx = \frac{\Gamma(n+1)}{a^{n+1}} for\hspace*{2pt} n > -1, Re(a) > 0$ [27] we get the expression for expectation as, 
\begin{equation}
\mathbb{E}[R_n]=\bigg( \frac{1}{\lambda v_m}\bigg)^{\frac{1}{m}}\frac{\Gamma \bigg(n+\frac{1}{m}\bigg)}{\Gamma(n)}
\end{equation}
\begin{figure}[!t]
	\centering
	\includegraphics[width=2.5 in]{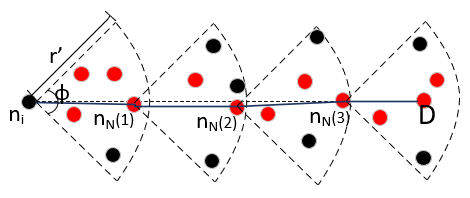}
	% where an .eps filename suffix will be assumed under latex, 
	% and a .pdf suffix will be assumed for pdflatex; or what has been declared
	% via \DeclareGraphicsExtensions.
	\caption{Each relay node forwards the packet to the  node satisfying coverage and collision constraints with least divergence from $n_i-D$ axis (direct path) and nearest to $D$ in a sector of angle $\phi$ $(0 \leq \phi \leq \pi)$; UAV nodes satisfying coverage and collision constraints are indicated in red.}
	%\label{fig_sim}
\end{figure}
The process of next node selection at every hop is repeated till $D$  is reached. This end to end opportunistic routing procedure is summarized in Algorithm 1. 

\begin{algorithm}
	\caption{Multi-hop Opportunistic 3D Routing Algorithm (MO3DR Algorithm)}
			Notations:\\
	$\mathbb{S}$: scanning sectoral area;\\
	$\phi$: scanning angle; \\
	$r^{'}$: scanning radius;\\
	%$P_{cov}$: Probability of Coverage;  \\
	%$P_{coll}$: Probability of Collision; \\
	$P_{cov}^*$: application specific probability of coverage; \\
	$P_{coll}^*$: application specific probability of collision; \\
	$r$: inter UAV distance;\\ 
	$R_{max}^*$: application specific maximum admissible $r$;  \\
	$R_{min}^*$: application specific minimum admissible $r$; \\	
	$\Psi_S$: set of SUAV;\\
	$\Psi_A$: set of AUAV\\
	$D$: destination node;\\
	$n_i$: $i^{th}$  SUAV node relaying the data;\\
	$n_j$: $j^{th}$  AUAV node;\\
	$D$: destination node;\\
	$n_N$: next hop node;\\
	$n_s$: set of nodes in sector $\mathbb{S}$;\\
	$n_p$: set of potential forwarder nodes in $\mathbb{S}$;\\
	%$n_*$: Node in $\mathbb{S}$ that is selected as $n_N$;\\
	\begin{algorithmic}[1] 
	\WHILE {$n_N \ne D$}
		\FOR{$n_i \in \Psi_S$}
		  \STATE find $| n_i-n_j|$ $\forall$  $n_j \in \Psi_A$
		  \STATE find $n_j \in \Psi_A$ with minimum $| n_i-n_j|$
		  \STATE update $ D \leftarrow n_j $ 
		  \STATE demarcate $\mathbb{S}$ along $ \pm \phi/2$ around $n_i-D$ axis
		  \STATE calculate $P_{cov}^* (8)$ and $P_{coll}^* (20)$
		  \STATE deduce $R_{max}^*$ and $R_{min}^*$
		    \FOR{$n_s \in\mathbb{S}$ }
	           \IF {$R_{min}^* \leq |n_i-n_s| \leq R_{max}^* \forall n_s \in \mathbb{S}$ }
		          \STATE update $n_p \leftarrow n_s$
		       \ELSE
		          \STATE ignore $n_s$
	           \ENDIF
	    	\ENDFOR
	    	\FOR {$n_p \in\mathbb{S}$}
	    	   \IF {$D \in n_p$}
                  \STATE update $n_N\leftarrow D$
               \ELSE
                   \STATE find $|n_i-n_p|$ and $n_i-n_p$ axis divergence from $n_i-D$ axis $\forall n_p \in \mathbb{S}$
                   \STATE find $n_p \in \mathbb{S}$ such that $|n_i-n_p|$  \&\& $ n_i-n_p$ axis divergence from $n_i-D$ axis = minimum
                   \STATE update $n_N \leftarrow n_p$
                   \STATE transmit the data packet to $n_N$
                   \STATE $n_i \leftarrow n_N$
               \ENDIF
             \ENDFOR 
		\ENDFOR
	\ENDWHILE	
	\end{algorithmic} 
\end{algorithm} 

As expressed in Algorithm 1, routing procedure is initiated when a SUAV $n_i \in \Psi_S$ is ready to relay the data packet. The Euclidean distance ($|n_i-n_j|$) between transmitting SUAV and the AUAVs $\forall$ $n_j \in \Psi_A$  is calculated. The nearest AUAV to $i^{th}$ transmitting SUAV is marked as the destination node $D$. The objective is to transmit the data packet from $n_i$ to $D$ via multi-hop path. To select the next hop nodes across the multi-hop path, sector $\mathbb{S}$ (with sector angle = $\phi$ and sector radius = $r'$) is demarcated along $\pm \phi/2$ around $n_i-D$ axis (direct path). The set of nodes within $\mathbb{S}$ are denoted as $n_s$. Amongst these nodes we determine set of potential receptor nodes $n_p \subset n_s$ that lie within the admissible range of inter UAV distance $R_{min}^* \leq r \leq R_{max}^*$  (deduced from  the coverage ($P_{cov}*$) and collision  ($P_{coll}*$) constraint of the application). We then select one particular node  $\in n_p$ as the next hop node $n_N$ to which the data packet is forwarded. If $D$ is one among $n_p$ it is selected as $n_N$ otherwise, the node nearest to $D$ and least divergent from the direct path is chosen as $n_N$. 
\section{Results and Discussion}
In this section, we present the results from analytical model and simulations. We have evaluated the performance of our analytical model by generating results based on the expression derived and validated them via simulations. The parameters used for simulation are stated in Table III.
\subsection{Probability of coverage $(P_{cov})$ and maximum operable inter UAV distance $(R_{max})$}
In this subsection we present the performance of UAV networks in terms of $P_{cov}$ for different values of inter UAV distance $r$, with changing network parameters such as pathloss component $\delta$, network density $\lambda$ and number of UAV nodes $N$. In figures 6, 7 and 8 $r$  and $P_{cov}$ are represented on x and y axis respectively. From Fig. 6 we observe that  $P_{cov}$ decreases with increase in $r$ and $\delta$. Because SIR at receptor UAV is inversely related to both the distance between itself and the transmitter UAV and path loss component (1). $P_{cov}$ being directly proportional to SIR; decrease in SIR causes corresponding decrease in $P_{cov}$. Thus, we see higher $P_{cov}$ when $\delta = 3.4$ as against $\delta = 4.2 $. 
\begin{table}
	\centering
	\caption {List of simulation parameters.}
	\begin{tabular} {l l}
		\hline
		%\hline\\
		\textbf{parameter} & \textbf{specification}\\
		\hline
		%\hline\\
		network simulator & MATLAB\\
		%\hline
		network area & 1000m x 1000m \\
		%\hline
		%Number of disaster location & 3-5\\
		%\hline
		number of UAVs & 5-100\\
		%\hline
		transmission range & dynamic\\
		%\hline
		altitude & 250-300m\\
		%\hline
		propagating model& Nakagami  small scale fading channel\\
		%\hline
		pathloss component & 3.4, 4.2\\
		%\hline
		%hline
		$\theta$ (SIR threshold) & 0 dB\\
		%\hline
		maximum number of iteration & 50\\
		\hline
	\end{tabular}
	%\caption {List of simulation parameters.}
\end{table}

In Fig. 7 we consider the network performance when density of UAV nodes ($\lambda$) are increased. We can notice that for greater $\lambda$, $P_{cov}$ degrades. This is due to the subsequent increase in the inter UAV interference with increase in $\lambda$. Though, the number of UAVs available for communication also increases with increase in $\lambda$, the impact  of inter UAV interference is not counteracted by the benefit gained from raise in the number of UAV nodes that are accessible for establishment of connection. Hence, we observe decline in $P_{cov}$ with increase in $\lambda$.
\begin{figure}[!t]
	\centering
	\includegraphics[width=2.5 in]{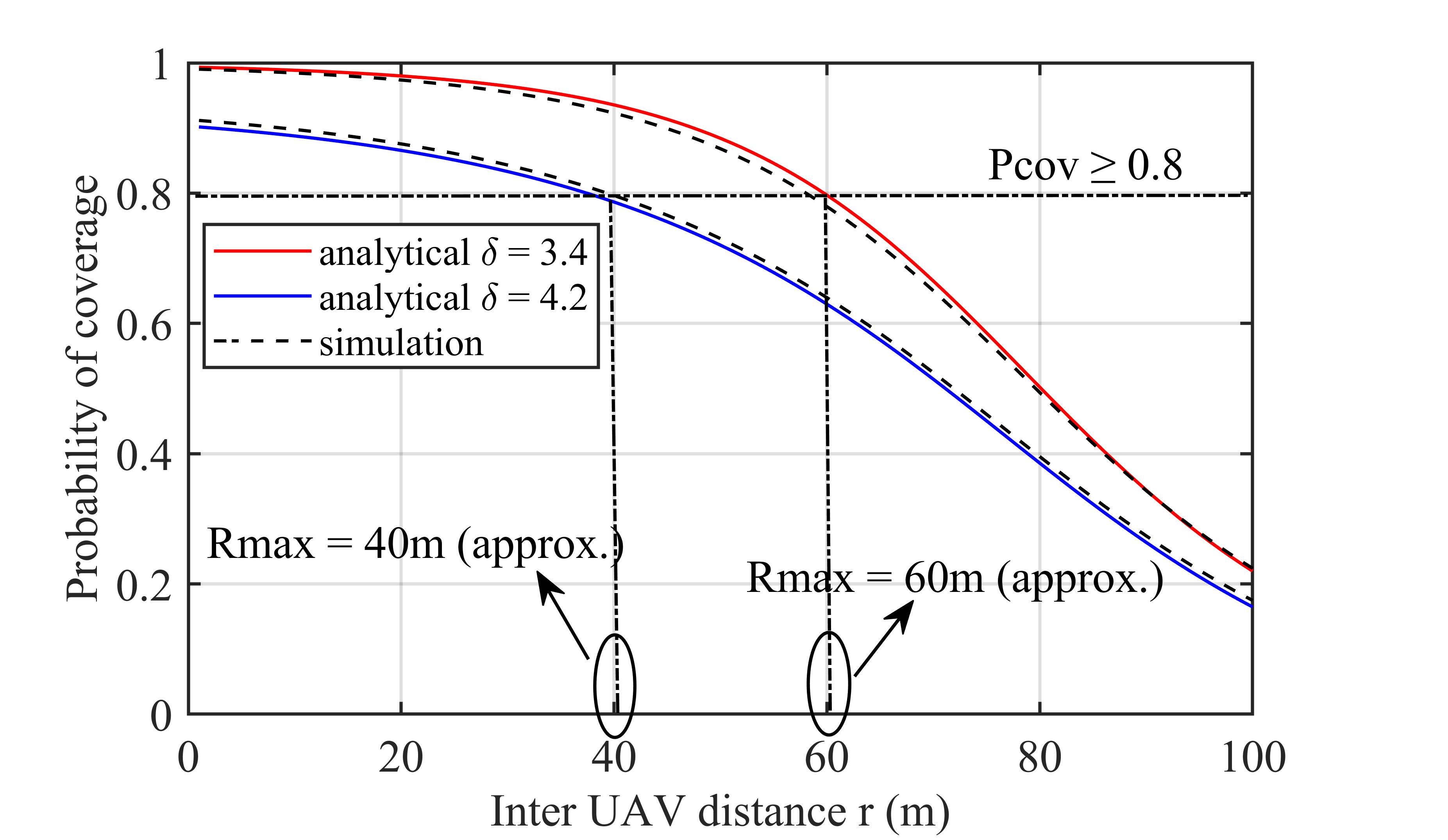}
	%	
	%	% where an .eps filename suffix will be assumed under latex, 
	%	% and a .pdf suffix will be assumed for pdflatex; or what has been declared
	%	% via \DeclareGraphicsExtensions.
	\caption{Probability of coverage ($P_{cov}$) as a function of inter UAV distance ($r$) for varying path loss component ($\delta$).}
	%	%\label{fig_sim}
\end{figure}

Fig. 8 represents $P_{cov}$ for varying number of UAV nodes. We can see that varying the number of UAV nodes will vary the inter UAV distance at which maximum $P_{cov}$ is achieved. Because increase in number of UAVs will increase the possibility of collision and inter UAV interference. Contrarily lesser the number of UAV nodes leads to sparsely deployed UAV network, thus, creating the problem of dearth of receptor UAVs within the coverage range of transmitting UAV. Hence, to counterbalance the above issues inter UAV distance  at which maximum $P_{cov}$ is attained differs with changing number of UAV nodes in the network. 

\begin{figure}[!t]
	\centering
	\includegraphics[width=2.5 in]{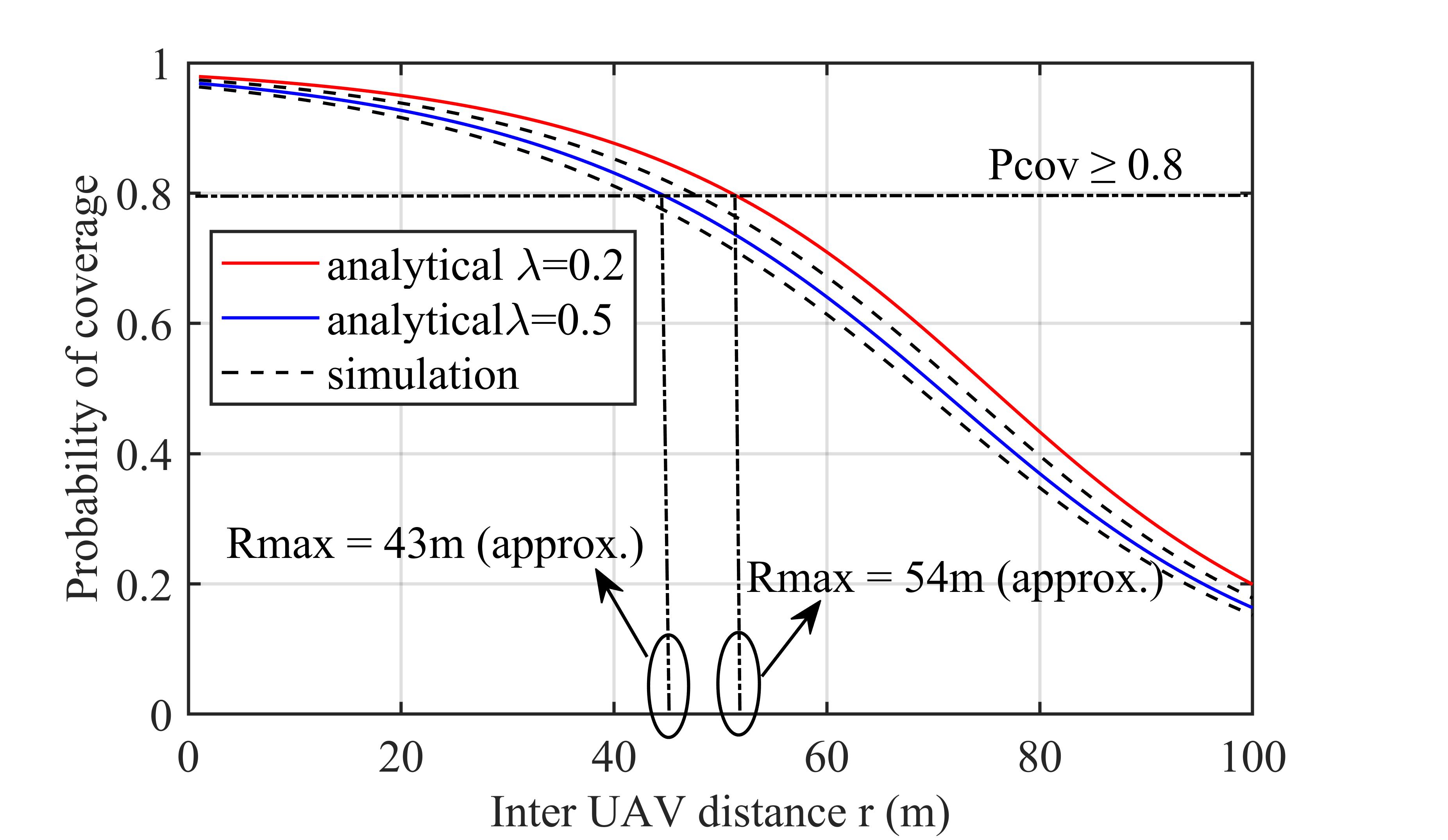}
	% where an .eps filename suffix will be assumed under latex, 
	% and a .pdf suffix will be assumed for pdflatex; or what has been declared
	% via \DeclareGraphicsExtensions.
	\caption{Probability of coverage ( $P_{cov}$) as a function of inter UAV distance ($r$) for varying network density ($\lambda$).}
	%\label{fig_sim}
\end{figure}

\begin{figure}[!t]
	\centering
	\includegraphics[width=2.5 in]{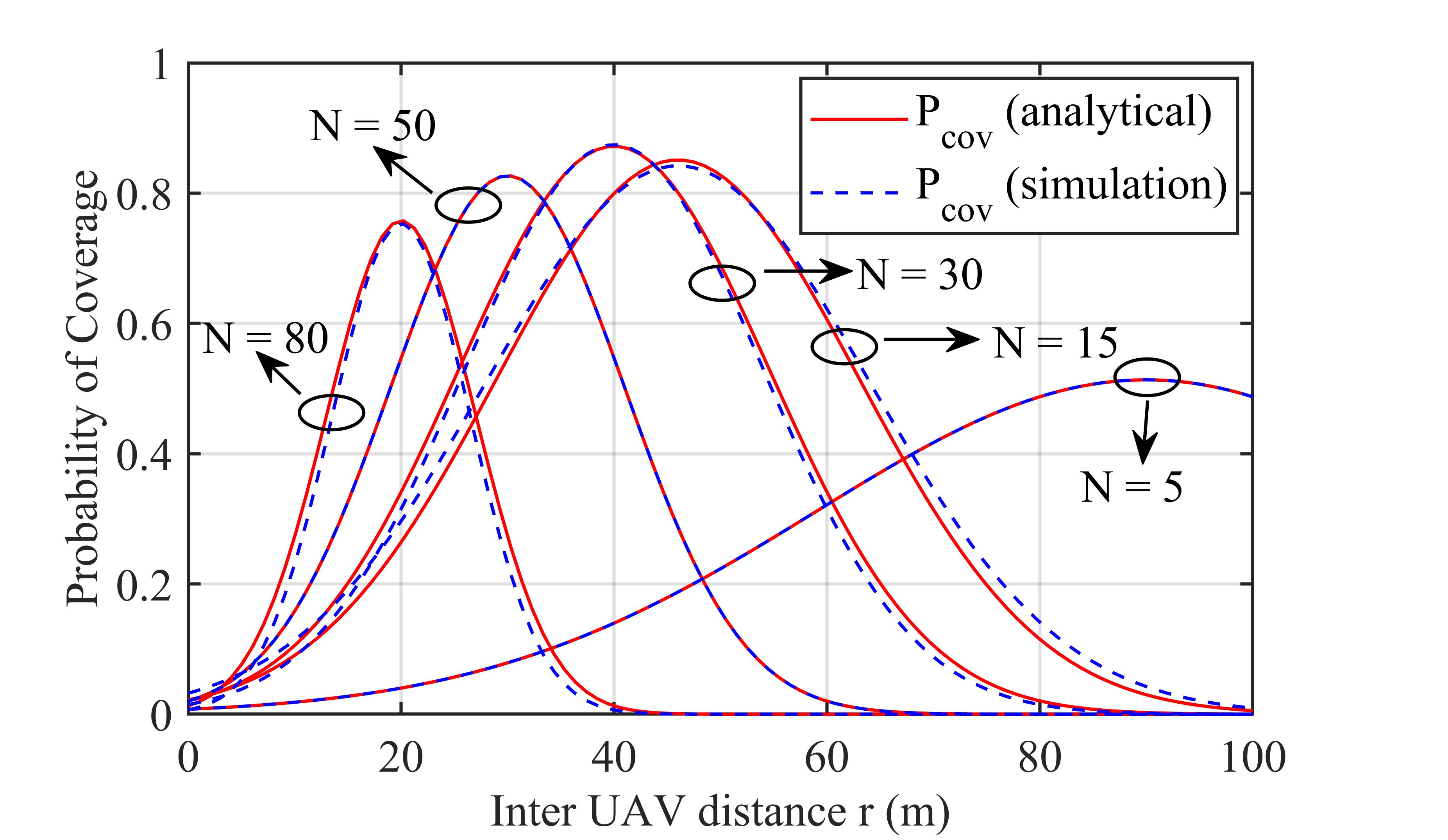}
	% where an .eps filename suffix will be assumed under latex, 
	% and a .pdf suffix will be assumed for pdflatex; or what has been declared
	% via \DeclareGraphicsExtensions.
	\caption{Probability of coverage ( $P_{cov}$) as a function of inter UAV distance ($r$) for varying number of UAV nodes ($N$).}
	%\label{fig_sim}
\end{figure}

Therefore, to ensure reliable communication between UAVs, connectivity is established only when receiver UAV is within the coverage range of transmitting UAV. Thus, restricting the maximum admissible inter UAV distance ($R_{max}$). If the application  requires $P_{cov} \geq 0.8$, we empirically deduce the corresponding $R_{max}$ that would satisfy $P_{cov}=0.8$ as depicted in figures 6 and 7. In Fig. 6 we  notice that $R_{max}$ is deduced to be equal to $40$m and $60$m (approx.) for $\delta=4.2$ and $\delta=3.4$ respectively. Similarly in Fig. 7 we obtain $R_{max}$ equal to $43$m and $54$m (approx.) for $\lambda=4.2$ and $\lambda=3.4$ respectively. 
\begin{figure}[!t]
	\centering
	\includegraphics[width=2.5 in]{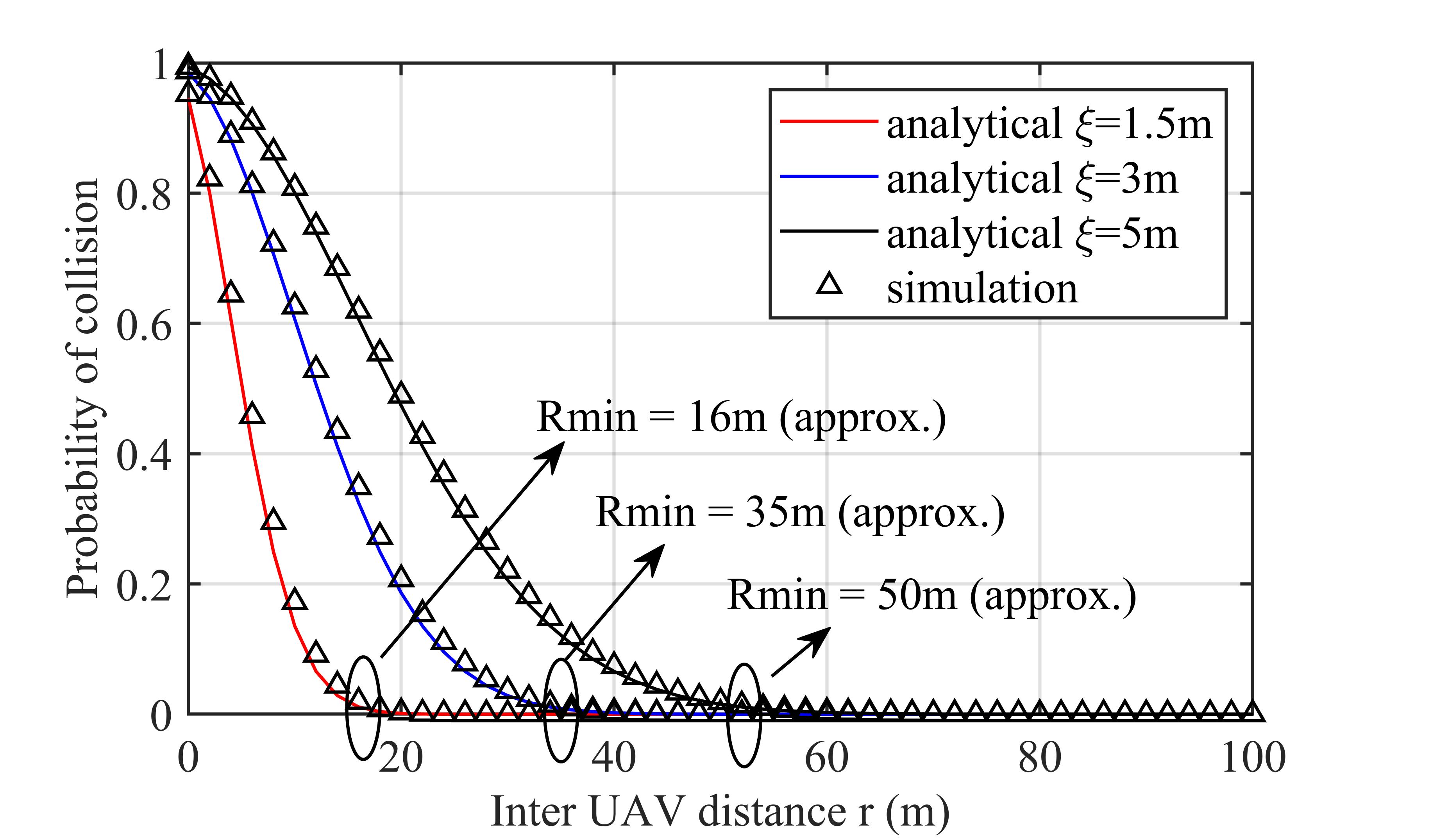}
	%\includegraphics[width=\linewidth,]{imagefile}
	% where an .eps filename suffix will be assumed under latex, 
	% and a .pdf suffix will be assumed for pdflatex; or what has been declared
	% via \DeclareGraphicsExtensions.
	\caption{Probability of collision ($P_{coll}$) as a function of  inter UAV distance ($r$) for varying trajectory divergence ($\xi$).}
	%\label{fig_sim}
\end{figure}
\subsection{Probability of collision $(P_{coll})$ and minimum operable inter UAV distance $(R_{min})$}
In this subsection we explore the performance of UAV network with respect to $P_{coll}$ for various values of inter UAV distance $r$ with changing trajectory divergence $\xi$. In Fig. 9 $r$  and $P_{coll}$ are  represented on x  and y axis respectively. From Fig. 9 it can be observed that  $P_{coll}$ increases with increase in $\xi$ and decrease in $r$. Higher the $\xi$ of UAV node , higher will be it's positional uncertainty hence, increasing the likelihood of collision. To make sure that two communicating UAVs do not collide, the link between them is set up only if they are at a safe distance from each other ($R_{min}$). From Fig. 9 we can infer that $R_{min} = 16$, $35$m and $50$m (approx.) for $\xi = 1.5$m, $3$m and $5$m respectively corresponding to $P_{coll} = 0$. 

Therefore, to enable disruption free communication between UAVs  the coverage and collision constraints have to be met. To meet the $P_{cov}$ and $P_{coll}$ requirements of the underlying application, communication is established only between those UAVs that satisfy the inter UAV distance condition $R_{min} \leq r \leq R_{max}$. 
\subsection{Inter UAV routing using MO3DR}
As discussed in the system  model, the proposed MO3DR algorithm selects the next hop node within a specific sector $\mathbb{S}$ with sectoral angle $\phi$ along the direct path. Therefore, it is important to understand the impact of $\phi$ on inter UAV distance $r$. Figures 10 and 11 depict the PDF and CDF of $r$ respectively on y axis and $\phi$ on x axis. From these figures it is evident that varying the sector angle will vary the  average inter UAV distance. As detailed in Fig. 10, for $\phi = \pi/8$, $r$ is  $15$m (approx.) while for $\phi = 7\pi/8$,  $r$ is $4$m (approx.). Similarly, from Fig 11. it can be seen that for $\phi = \pi/8$, $r$ is $16$m (approx.) whereas for $\phi = 7\pi/8$, $r$ is $5$m (approx.).
\begin{figure}[!t]
	\centering
	\includegraphics[width=2.5 in]{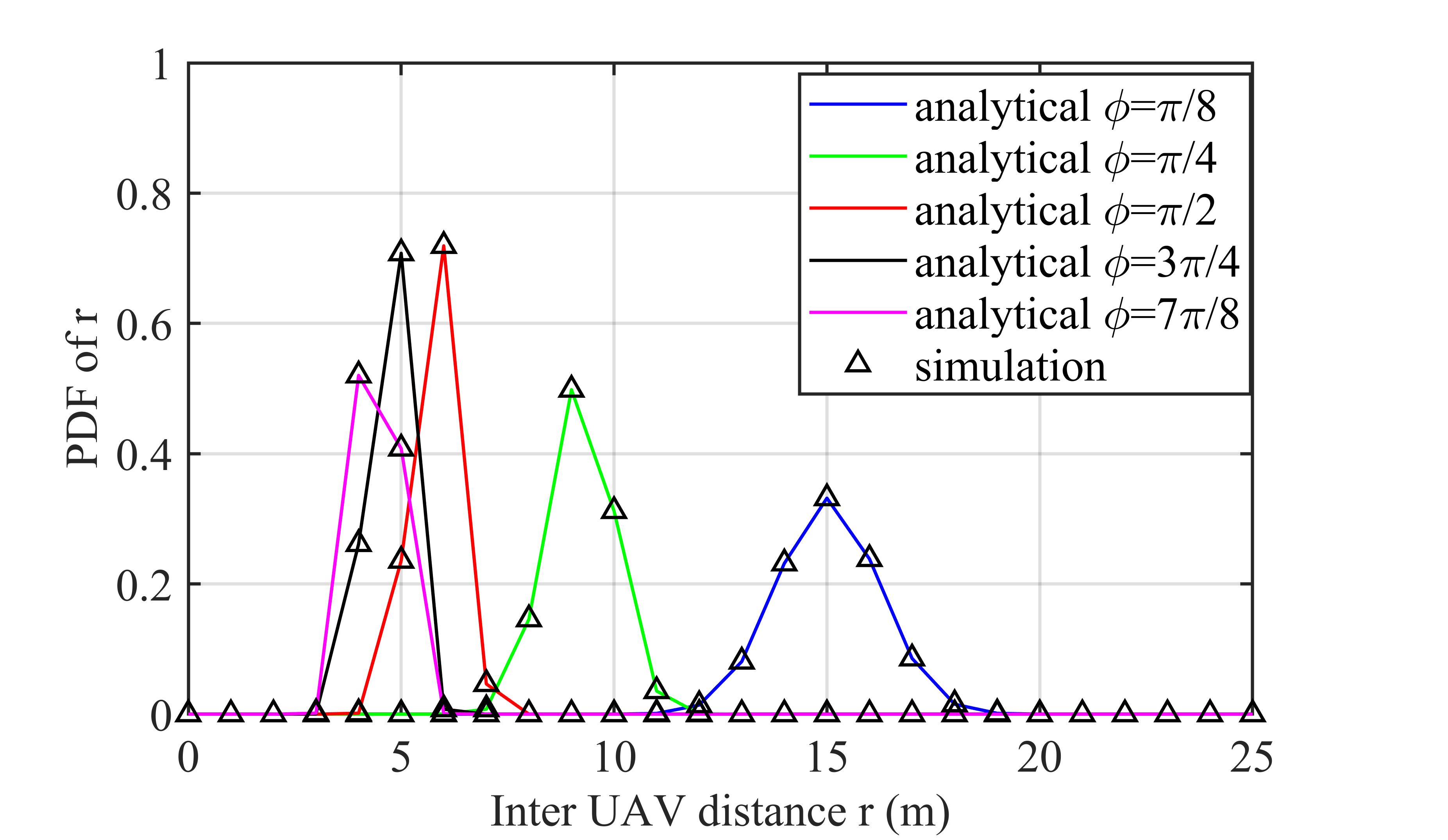}
	% where an .eps filename suffix will be assumed under latex, 
	% and a .pdf suffix will be assumed for pdflatex; or what has been declared
	% via \DeclareGraphicsExtensions.
	\caption{Probability distribution function ($f_{R_n}$) of inter UAV distance ($r$)  for varying sector angle ($\phi$).}
	%\label{fig_sim}
\end{figure}
This signifies that greater $r$ is achieved  at smaller $\phi$.  Because, smaller the $\phi$ smaller is the divergence of UAV nodes within $\mathbb{S}$ from the direct path, thus allowing larger $r$. Conversely, larger the $\phi$, larger could be the divergence of UAV nodes within  $\mathbb{S}$ from the direct path hence, decreasing $r$. The parameter $r$ has direct impact on $\mathbb{E}[R_n]$, farther the receptor node from the transmitting node greater is the $\mathbb{E}[R_n]$ and vice-versa assuming divergence from the direct path to be minimal. Therefore, it can be concluded that $\phi$ has significant impact on $\mathbb{E}[R_n]$.  
\begin{figure}[!t]	\centering
	\includegraphics[width=2.5 in]{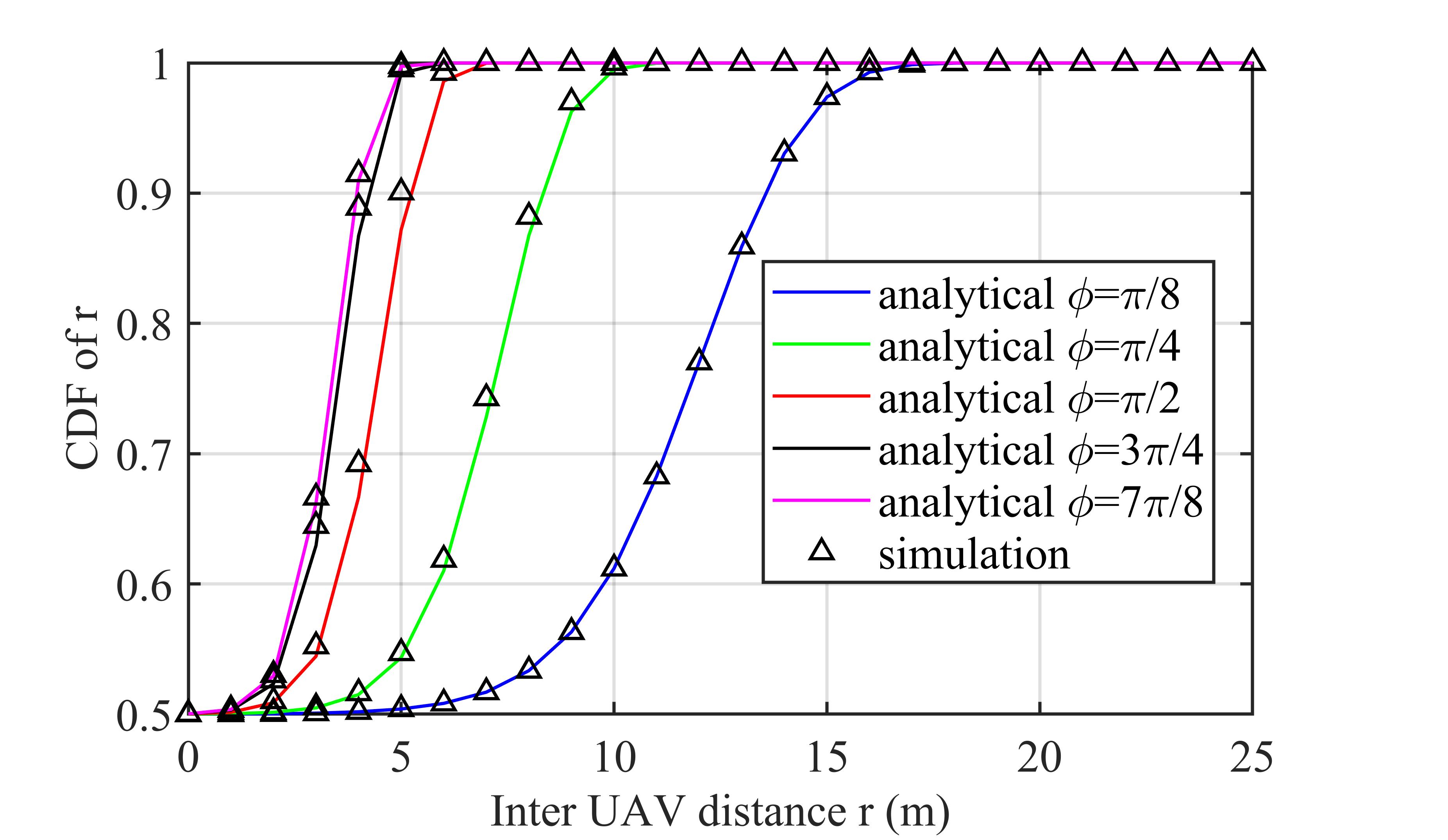}
	% where an .eps filename suffix will be assumed under latex, 
	% and a .pdf suffix will be assumed for pdflatex; or what has been declared
	% via \DeclareGraphicsExtensions.
	\caption{Cumulative distribution function ($F_{R_n}$) of inter UAV distance $(r)$ for varying sector angle ($\phi$).}
	%\label{fig_sim}
\end{figure}

In Fig. 12 sectoral angle is represented on x axis and expected progress towards the destination is represented on y axis. From Fig. 12 we can infer that as $\phi \to 0$ or $\phi \to 2\pi$, $\mathbb{E}[R_n]$ decreases. Firstly, as $\phi \to 0$ there could be very few or no potential receptor nodes  within  $\mathbb{S}$. Secondly, larger the $\phi$, larger could be the divergence of potential receptor nodes from the direct path. The graph indicates that for $\pi/2 < \phi < 3\pi/4$, $\mathbb{E}[R_n]$ is maximum for varying values of $N$.
\begin{figure}[!t]	\centering
	\includegraphics[width=2.5 in]{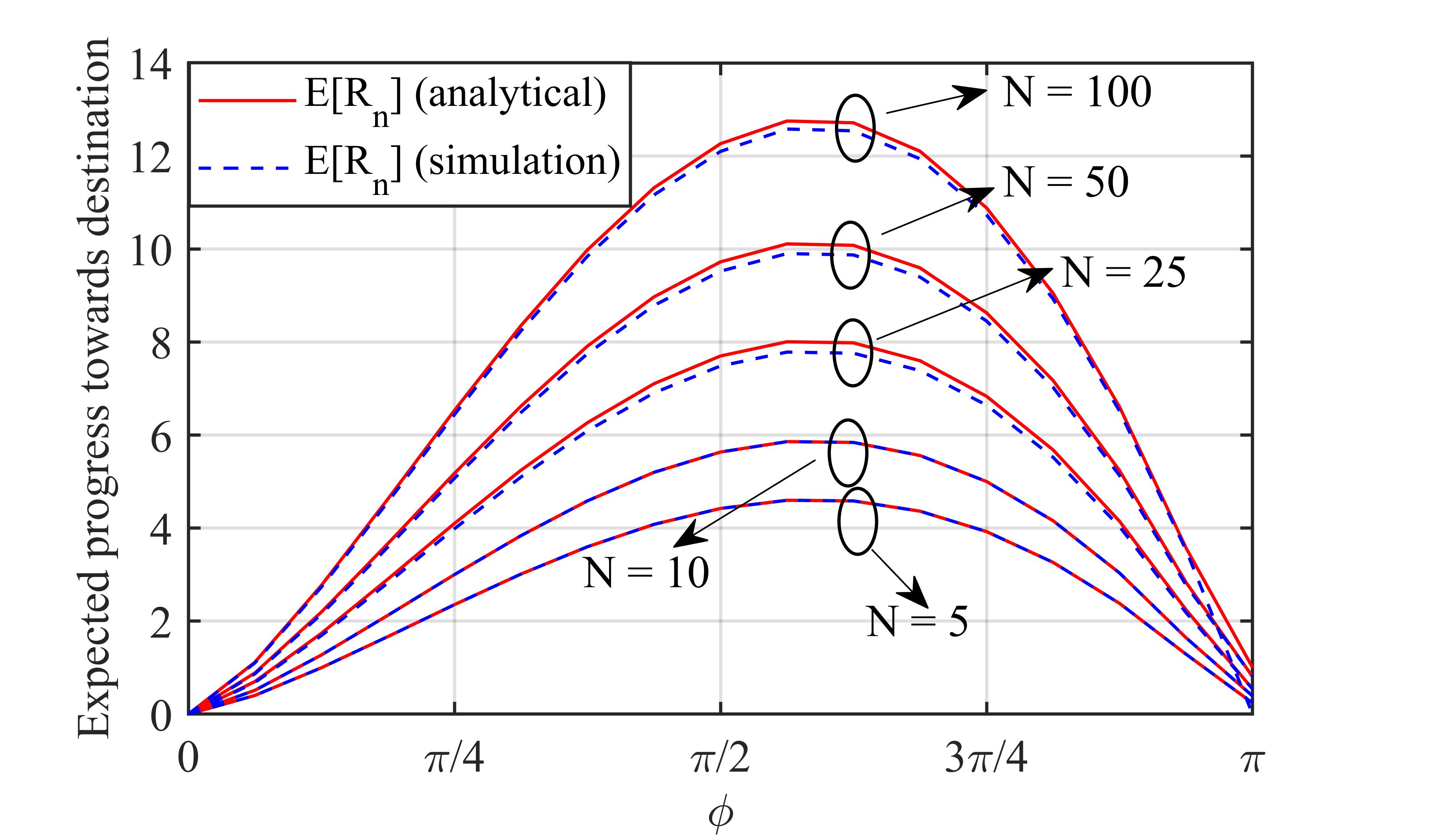}
	% where an .eps filename suffix will be assumed under latex, 
	% and a .pdf suffix will be assumed for pdflatex; or what has been declared
	% via \DeclareGraphicsExtensions.
	\caption{The expected progress of packets towards destination ($\mathbb{E}[R_n]$)  as a function of sector angle ($\phi$) for varying  number of UAV nodes ($N$).}
	%\label{fig_sim}
\end{figure}

\begin{figure}[!t]
	\centering
	\includegraphics[width=2.5 in]{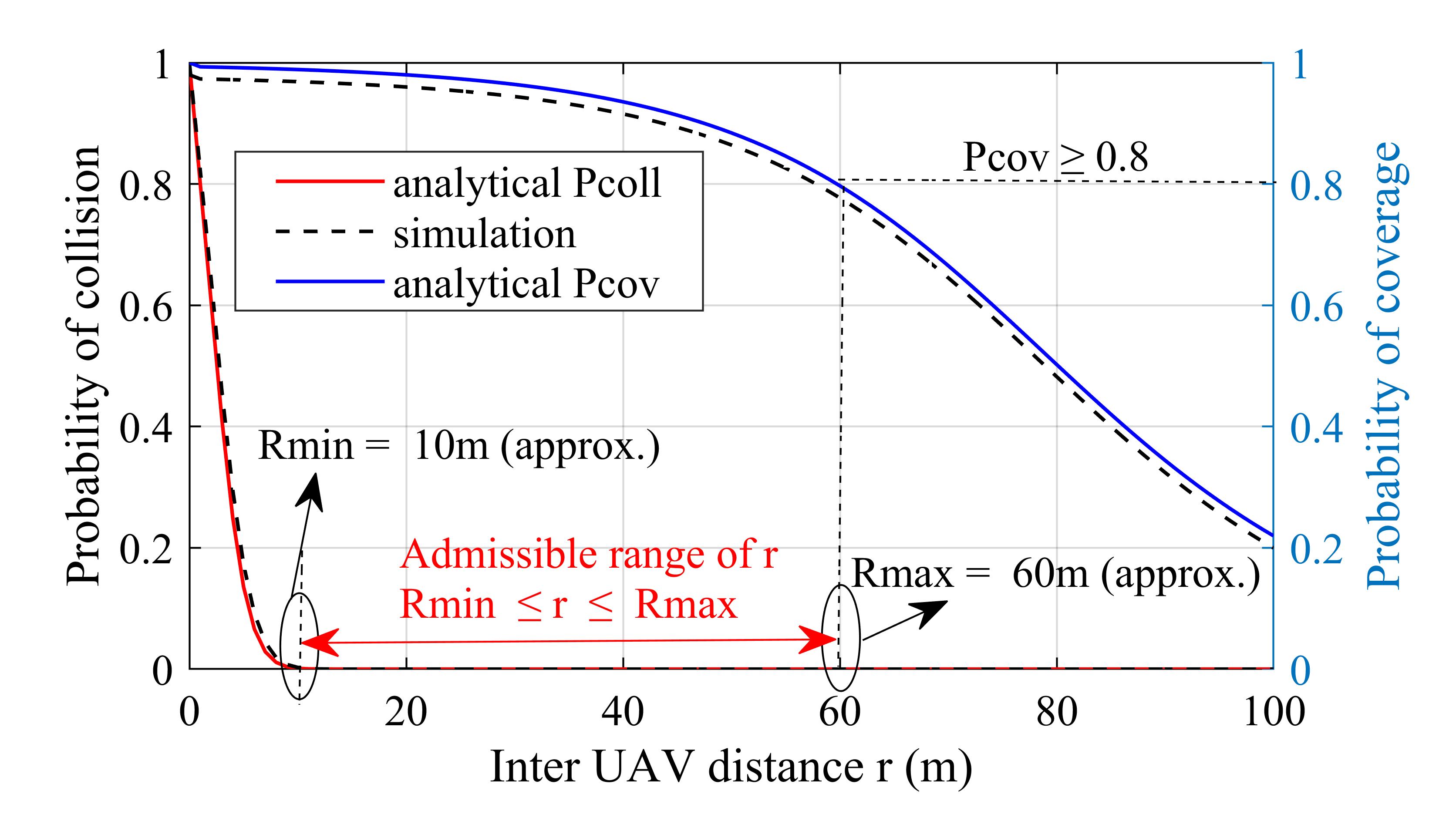}
	% where an .eps filename suffix will be assumed under latex, 
	% and a .pdf suffix will be assumed for pdflatex; or what has been declared
	% via \DeclareGraphicsExtensions.
	\caption{Probability of collision ($P_{coll}$) and Probability of coverage ($P_{cov}$)  as a function of inter UAV distance ($r$).}
	%\label{fig_sim}
\end{figure}
\subsection{Empirical optimality condition for inter UAV distance ($R_{min}$ and $R_{max}$) and sector angle $\phi$ }
In this subsection we analyse the proposed system end to end for a specific use case of $N=25$, $\delta = 3.4$, $\lambda = 0.2$ and $\xi = 0.8$m. First, we empirically deduce admissible range of inter UAV distance ($R_{min} \leq r\leq R_{max}$) from the coverage and collision constraints as per the application demand. Then, for the given admissible range of $r$ we study the behaviour of   $\mathbb{E}[R_n]$ for varying $\phi$. In Fig. 13 $r$ is represented on x axis, $P_{coll}$ and $P_{cov}$ on left and right y axis respectively. We obtain admissible range of inter UAV distance to  be equal to $10$m $\leq r\leq$ $60$m conforming to the application requirements of  $P_{coll} = 0$ and $P_{cov} \geq 0.8$. Fig. 14 elucidates $R_{min} \leq r\leq R_{max}$ on x axis and $\mathbb{E}[R_n]$ on y axis. $\mathbb{E}[R_n]$ increases with increase in $r$ for $\phi$ varying from $\pi/4$ to $3\pi/4$. Higher $\mathbb{E}[R_n]$ is observed for $\pi/2 \leq \phi \leq 3\pi/4$. For $\phi = \pi, \mathbb{E}[R_n]$ decreases for larger values of $r$. Because as $\phi$ increases, divergence from the direct path increases, thus, decreasing $\mathbb{E}[R_n]$  for higher values of $r$.  Hence, it can be inferred that for the assumed system parameters  $N=25$, $\delta = 3.4$, $\lambda = 0.2$, $\xi = 0.8$m with application requirement of $P_{coll} = 0$ and $P_{cov} = 0.8$, choosing $\phi = 2\pi/3$ would maximise $\mathbb{E}[R_n]$ for the range  $10$m $\leq r\leq$ $60$m. 
\begin{figure}[!t]
	\centering
	\includegraphics[width=2.5 in]{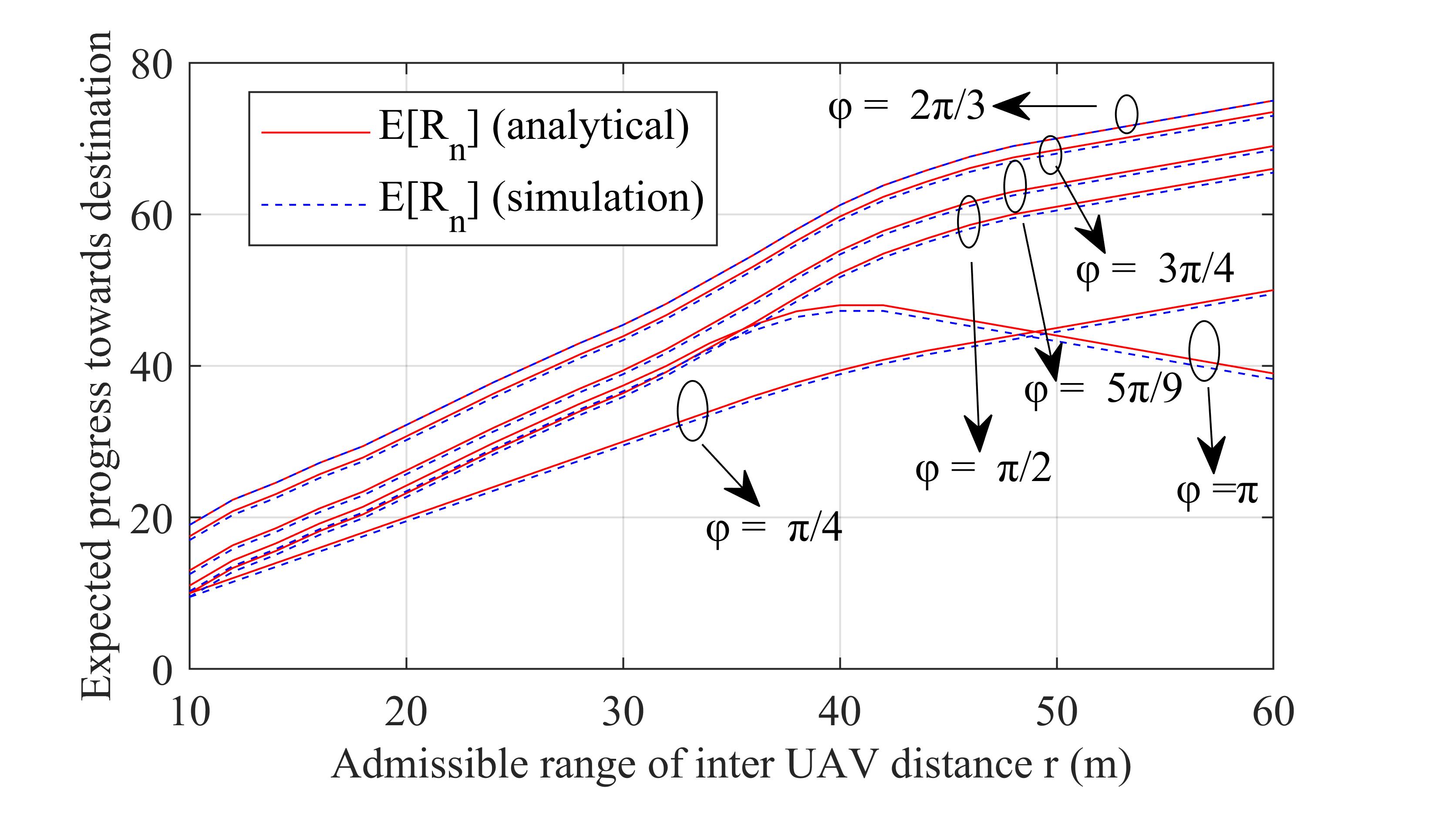}
	% where an .eps filename suffix will be assumed under latex, 
	% and a .pdf suffix will be assumed for pdflatex; or what has been declared
	% via \DeclareGraphicsExtensions.
	\caption{The expected progress of packets towards destination ($\mathbb{E}[R_n]$)  for admissible range of inter UAV distance ($R_{min} \leq r \leq R_{max}$)  with varying sector angle ($\phi$).}
	%\label{fig_sim}
\end{figure}

 \section{Conclusion}
 In this paper, we study the downlink performance of UAV network for disaster surveillance application. We introduce a stochastic geometry based approach to model the 3D UAV network deployed for emergency communication, considering various network parameters such as LoS Nakagami $m$ small scale fading channel, UAV network density, SIR, trajectory divergence and Poisson distributed inter UAV distance. First, we derive closed form mathematical expression for $P_{cov}$. Then we analyse the performance of UAV network in terms of $P_{cov}$ for different values of $r$. Second, we formulate the analytical expression for $P_{coll}$ and then further explore the behaviour of the UAV network with regard to $P_{coll}$ for changing values of $r$.  Third, we empirically deduce the operable range of inter UAV distance $R_{min} \leq r \leq R_{max}$ such that reliable communication is established by satisfying the coverage and collision constraints of the  underlying application. Lastly, we propose a novel MO3DR algorithm for inter UAV routing such that at each hop data packet makes maximum progress towards the destination node considering inter UAV coverage and collision constraints. The results show that,  the network parameters such as pathloss component, network density and the number of UAV nodes have significant impact on  $P_{cov}$, whereas trajectory divergence has profound influence on $P_{coll}$. The numerical results obtained using analytically derived closed form mathematical expressions are verified using simulations. Further extensions to this framework may include incorporating high density UAV network with dynamically changing AUAV positions. 

\section*{Acknowledgement}
The authors thank the Department of Science and Technology (DST), Government of India, for funding this work under the project Advanced Communication System included in  National Mission on Interdisciplinary Cyber Physical Systems. 
\appendices
\section{}
In this appendix we derive the Laplace transform $\mathcal{L}_I(s)$. We define aggregate interference power as,
%\begin{flalign}\tag{21}
%&I=\sum_{i\in \psi_S\x}^{}\mathcal{H}_{ij}l(r_i,h_i)^{\frac{-\alpha}{2}}&
%\end{flalign}
\begin{equation}
\mathcal{L}_I(s)= \mathbb{E}_I[e^{-sI}]
\end{equation}
expanding the above equation we get,
\begin{equation}
\mathcal{L}_I(s)= \mathbb{E}_{\Psi_S}\bigg[\prod_{i \in \Psi_s \setminus \{x\}}\bigg [ \mathbb{E}_{\mathcal{H}_{ij}} \bigg[exp-(s_{\mathcal{H}_{ij}}l(r_{ij},h_i,\alpha))\bigg]\bigg] \bigg]
\end{equation}
where $\{x\}$ corresponds to transmitting UAV, for which any other transmitting UAV $i\in\Psi_s  \setminus \{x\}$ acts as interference. By applying moment generating function of the Gamma and exponential distributions to the  expectation term $\mathbb{E}_{\mathcal{H}_{ij}}[\dots]$ we get (32).
\begin{equation}
\mathcal{L}_I(s)=\mathbb{E}_{\Psi_S}\bigg[\prod_{i \in \Psi_s \setminus \{x\}} \frac{1}{(1+\frac{s}{\delta}l(r_{ij},h_i,\alpha))^\delta}\bigg]
\end{equation}
By using probability generating functional of m-dimensional PPP we get the final expression as,
\begin{equation}
\mathcal{L}_I(s)=exp\bigg[ -2\pi\lambda\int_{v_mr^m} 1- \frac{1}{(1+\frac{s}{\delta}l(t, \alpha))^\delta}tdt\bigg]
\end{equation}

% use section* for acknowledgement
%\section*{Acknowledgment}
%
%
%The authors would like to thank...

% Can use something like this to put references on a page
% by themselves when using endfloat and the captionsoff option.
\ifCLASSOPTIONcaptionsoff
  \newpage
\fi

\end{document}